\documentstyle[aps,floats,twocolumn,prl]{revtex}

\def\beq{\begin{equation}}
\def\eeq{\end{equation}}
\def\beqa{\begin{eqnarray}}
\def\eeqa{\end{eqnarray}}
\newcommand{\beqas}{\begin{eqnarray*}}
\newcommand{\eeqas}{\end{eqnarray*}}

\def\tr{\mathop{\rm Tr}}

\def\oneh{\frac{1}{2}}
\def\D{\Delta}
\def\mb{{\bf m}}
\def\chib{\mbox{\boldmath $\chi$}}
\def\gab{\mbox{\boldmath $\gamma$}}

\def\Sb{{\bf S}}
\def\hS{\hat{S}}
\def\hSb{{\bf \hat{S}}}

\def\Jb{{\bf J}}
\def\mb{{\bf m}}
\def\Mb{{\bf M}}

\def\Ab{{\bf A}}
\def\Bb{{\bf B}}

\def\Lab{{\bf \Lambda}}

\def\la{\lambda}

\def\t{\theta}
\def\tb{\mbox{\boldmath $\t$}}
\def\lab{\mbox{\boldmath $\la$}}

\def\<{\langle}
\def\>{\rangle}

\newcommand{\cut}[1]{}

\begin{document}
\title{\bf Tractable approximations for
probabilistic models:\\ The adaptive TAP mean field approach}

\author{Manfred Opper$^1$ and Ole Winther$^{2,3}$} 

\address{$^1$Neural Computing Research Group,
School of Engineering and Applied Science,
Aston University,
Birmingham B4 7ET, United Kingdom\\
$^2$Theoretical Physics II, Lund University, S\"{o}lvegatan 14 A, 
S-223 62 Lund, Sweden\\
$^3$Department of Mathematical Modelling, 
Technical University of Denmark B321,
DK-2800 Lyngby, Denmark}
\maketitle
\begin{abstract}
We develop an advanced mean field method for approximating averages in
probabilistic data models that is based on the TAP approach 
of disorder physics. In contrast to conventional TAP,
where the knowledge of the distribution of couplings between 
the random variables is required, our method adapts to the concrete 
couplings. We demonstrate the validity of our approach, which is
sofar restricted to models with non-glassy behaviour, by replica 
calculations for a wide class of models as well as by simulations 
for a real data set.
\end{abstract}
Probabilistic models (for a review see e.g.\ \cite{Jo99})
\iffalse
probabilistic independent component analysis \cite{HoWiHa01} (ICA),
or Gaussian process models CITE WILLIAMS 
\fi
find widespread applications in many areas 
of data modeling. Their goal is to explain complex observed data
by a set of unobserved, {\em hidden} random variables based 
on the joint distribution of both sets of variables. 
The price that a modeler has to pay 
for the high degree of flexibility of these models is the vast increase in 
computational complexity when the number of hidden variables is large.

Both statistical inference about hidden variables and training usually 
require computation of marginal distributions of the hidden variables 
which for exact calculation demands infeasible
high dimensional sums or integrals.
Since similar types of calculations are ubiquitous in the computations
of thermal averages, there is a great deal of interest
in adopting approximation techniques from statistical physics. 
For a variety of cases, when a standard tool, the Monte Carlo
sampling technique reaches its limits, a simple  
mean field (MF) method, 
which neglects correlations of random variables has yielded good 
results in a variety of probabilistic data models. 
The MF approximation yields a closed set of nonlinear
equations for the approximate expectation values of random variables
which usually can be solved in a time that only grows polynomially 
in the number of variables. At present, there is a growing research 
activity trying to overcome the limitations of the simple MF method
by partly including the dependencies of variables but
still keeping the approximation tractable (for a review see \cite{OpSa2001}). 

Various researchers
\cite{OpWi00,MePa86,MePaVi87,Me89,Wo95,OpWi96,KaSa98,KaRo99,Ta98,BaKe00} 
have discussed applications of the so-called TAP MF approach,
originating in the statistical physics of
disordered systems, first introduced by Thouless, Anderson
and Palmer (TAP) \cite{ThAnPa77} to treat the Sherrington-Kirkpatrick (SK) model of
disordered magnetic materials \cite{SK75}.
Under the assumption that the couplings (interactions)
between random variables are themselves drawn at random from certain
classes of distributions, the TAP equations become {\em exact} 
in the thermodynamic limit of infinitely many variables.  
Unfortunately, the {\em Onsager correction} to the simple, naive
MF theory will explicitly depend on the distribution of these
couplings. Two models with the same connectivities but different
distributions for the couplings, like e.g.\
the SK model and the Hopfield model \cite{Hop82}
have different expressions for the Onsager corrections
(see e.g. \cite{MePaVi87}, chapter XIII).

In order to use the TAP method as a good approximation 
for models of real data, the lack of knowledge of the underlying
distribution of the couplings (which are usually functions of
the observed data) should be
compensated by an algorithm which {\em adapts} the
Onsager correction to the {\em concrete} set of couplings.
Simply taking the correction from a theory that {\em assumes}
a specific distribution may lead to suboptimal performance.        
This letter presents a solution to this problem for an
important class of probabilistic models. 
As a check of the validity of the approach, 
we show that our method leads to the exact results in the
thermodynamic limit for large classes of probability distributions over 
the couplings.
\iffalse
Our results also solve a longstanding problem in the 
statistical mechanics research on learning in neural networks 
(for a review, see e.g. \cite{SeSoTi92,Waal93,OpKi96}) by 
providing a general type of algorithms that is able able to achieve
the optimal generalization performance in large networks. 
\fi

We will consider probabilistic models of the type
\beq\label{model}
P(\Sb) = \frac{\rho(\Sb)}{Z(\tb,\Jb)}
\exp\left[\sum_{i<j} S_i J_{ij} S_j + \sum_i S_i \t_i\right]
\eeq
where the set $\Sb =(S_1,\ldots,S_N)$ denotes the (hidden) random variables
of the model. Any observed (i.e. fixed) quantities are assumed to be
encoded in the matrix $\Jb$ and the fields $\tb$. The term 
$\rho(\Sb)\equiv \prod_j \rho_j (S_j)$ is a product distribution which 
also contains all constraints of the $S_i$ 
(the range, discreteness, etc). 
In its simplest version, when $\Sb$ is a real variable
with positive measure $\rho$, the class of  
models (\ref{model}) contains Ising models (such the SK and Hopfield models), 
Gaussian process models \cite{OpWi00}, probabilistic independent
component analysis \cite{HoWiHa01} and
\iffalse 
and (by a suitable extensions to
vectorial variables and limit procedures)
\fi 
combinatorial 
optimization problems
\iffalse
, like the matching problem and the traveling salesman
problem 
\fi 
\cite{MePaVi87}. 
If we lift the restrictions that all variables must be real random variables,
we can treat a variety of important models with 
dependencies between the $S_i$ that are
defined through a set of fields $\sum_{i=1}^N x_{ij} S_i$.
We will give two examples. Bayesian learning in single layer neural networks
is described by a Gibbs distribution
$
P(\Sb)\propto P_0(\Sb) 
\prod_{j=1}^m F(\sum_{i=1}^N x_{ij} S_i),
$
where $\Sb$ is a weight vector of the network
being trained on a number of $m$ data vectors with components
$x_{ij}$ in a $N$ dimensional space.
$P_0$ is a prior distribution of the weights and $F$ is the Likelihood
quantifying the goodness of fit to the data \cite{OpWi96}.
A second example is given by the class of Bayesian belief networks
on a directed graph which are promising models for adaptive
expert systems. They are defined by
$
P(\Sb)=\prod_i P(S_i|\mbox{pa} (S_i))
$
where $S_i\in\{0,1\}$ and $pa$ denotes the parents of $S_i$, i.e.\
the variables in the graph that feed their information 
into $S_i$ via directed bonds. A specific type is the sigmoid belief networks
\cite{Ne92}, where 
$P(S_i|\mbox{pa} (S_i))=\frac{e^{S_i h_i}}{1+e^{h_i}}$ with 
$h_i =\sum_{j\in {pa}(S_i)} x_{ij} S_j$.
The latter two models can be easily brought into the form (\ref{model})
by the standard `field-theoretic' trick of introducing Dirac 
$\delta$-functions and their exponential representations using
purely imaginary conjugate variables $\hSb=(\hS_1,\ldots,\hS_m)$. 
This leads to an augmentation of the space of variables to the set 
$(\Sb,\hSb)$. The hatted variables
have the complex single variable distributions 
$
\hat{\rho}(\hS) = \int \frac{dh}{2\pi i} e^{-\hS h} e^{-H(h)} 
$
in case of the neural network model and
$
\hat{\rho}(\hS) = \int \frac{dh}{2\pi i} e^{-\hS h }/(1+e^h) 
$
for the belief network (where $m=N$).
The augmented coupling matrix is of the form
$
\Jb = \left(
\begin{array}{cc}
  \Ab & \Bb \\
  \Bb^T  & {\bf 0}
\end{array}
\right) \ ,
$ 
where $B_{ij} = x_{ij}$ and $\Ab=0$ for the neural 
network and $A_{ij}=B_{ij}=x_{ij}$ for
the belief net.

We will derive both an adaptive TAP-like approximation for the 
marginal distribution $P_i(S)\equiv \int \prod_{j\neq i} dS_j P(\Sb)$ and
the free energy $F(\Jb,\tb)= -\ln Z(\Jb,\tb)$.  
The free energy corresponds to 
the negative log probability of the observed data which can be
used as a yardstick for deciding which model best fits to the data.

Our derivation will be based on the cavity approach introduced by 
\cite{MePaVi87}. 
We will assume that we are not dealing with a glassy system with its
many ergodic components, but that all averages
are for a single state. This is (as shown for many of the teacher-student 
scenarios studied in the statistical mechanics of neural networks) 
usually expected to hold when the probabilistic model is well matched to
the data.
Defining the field $h_i = \sum_j J_{ij} S_j$, the marginal distribution 
of $S_i$ can be written as
\beqa\label{marginal}
P_i(S)= \int \prod_{j\neq i} dS_j P(\Sb)
= \frac{\rho_i(S)}{Z_i} e^{- H_i(S)} \ ,  
\eeqa
where we have introduced an effective single variable Hamiltonian $H_i(S)$
with corresponding partition function $Z_i$.
Defining an auxiliary average over the distribution of 
the system with variable $S_i$ left out by 
$\left\<\ldots\right\>_{\backslash i}$, we get 
\beq
- H_i(S)=  \ln{\left\<e^{S h_i}\right\>_{\backslash i}} = \sum_k 
\frac{\kappa_k^{(i)}}{k!} S^k
\eeq
where $\kappa_k^{(i)}$ are the cumulants of this {\em cavity} distribution,
i.e. $\kappa_1^{(i)} =\< h_i\>_{\backslash i}$ and
$\kappa_2^{(i)} = \< h_i^2\>_{\backslash i} - \< h_i\>_{\backslash i}^2$ etc.

The basic physical assumption, which is the major
ingredient of all cavity derivations of the TAP mean field theory \cite{MePaVi87},
is that all variables $S_j$ have only weak mutual dependencies.
Mathematically expressed within the so--called clustering hypothesis 
\cite{MePaVi87} this becomes equivalent to the {\em vanishing}
of all cumulants $\kappa_k^{(i)}$ with $k>2$ for fully connected
systems. In the case, where the $S_j$ are real variables with 
positive measure, this 
corresponds to a central limit theorem for the cavity fields.
Under this assumption, setting $V_i = \kappa_2^{(i)}$, we get 
\beqa
\label{cavfield}
\< h_i \> & = & \frac{1}{Z_i} 
\int dS \rho_i(S) \frac{\partial}{\partial S}
e^{-H_i(S)} = \<h_i \>_{\backslash i} + V_i  \<S_i\> \\
\label{marginal2}
P_i(S) & = & \frac{\rho_i(S)}{Z_i}
e^{\left(\sum_j J_{ij}\<S_j\> -  V_i \< S_i \> +\theta_i\right)S + \oneh V_i S^2}
%\exp\left[(\sum_j J_{ij}\<S_j\> -  V_i \< S_i \> +\theta_i)S + \oneh V_i S^2\right]
\eeqa
for $i=1,\ldots,N$. 
So far, the approach is well known. The new aspect of our paper is in 
the way we compute the $V_i$'s. Since these reaction terms 
account for the weak influence between random variables, they can be
computed self-consistently from the matrix of susceptibilities
$
\chi_{ij}\equiv\frac{\partial\<S_i\>}{\partial \t_j}
$.
We make the approximation 
that upon differentiation, the $V_i$'s are held constant
which is consistent with the fact that the $V_i$'s are expected to be
selfaveraging quantities in the thermodynamic limit. Under this 
assumption we get from eq.\ (\ref{marginal2})
$
\chi_{ij} = \chi_{ii} 
\left(\delta_{ij} + \sum_{k} (J_{ik} - V_k \delta_{ik}) 
\chi_{kj} \right)   
$
which can be solved with respect to $\chib$ and yields
$
\chib= (\Lab-\Jb)^{-1} \ ,
$
where $\Lab=\mbox{diag}\{V_i+1/\chi_{ii}\}$ is a diagonal matrix.
The {\em Fluctuation
Dissipation Theorem} (again assuming that we 
deal with a single state), shows that $\chib$ also equals the matrix of 
correlations $C_{ij} = \<S_j S_k\> -
\<S_j\>\<S_k\>$. By specializing to the diagonal elements, 
we can compute $V_i$ as a function
of $\<S_i^2\> - \<S_i\>^2$ by solving 
\beq\label{TAP:Var}
\<S_i^2\> - \<S_i\>^2 = \frac{\partial^2 \ln Z_i}{\partial \t_i^2}
=
\left[\left(\Lab-\Jb\right)^{-1}\right]_{ii}
\eeq
for $i=1,\ldots,N$.
The sets of equations (\ref{marginal2}) together with (\ref{TAP:Var}) 
constitute the first main result of this letter.
They yield closed sets of equations for the first and second
moments of $S_i$ which in turn 
enables us to approximate the full marginal distribution of $S_i$
and the correlation functions. For comparison we note that 
the naive mean field approximation (for real random variables) 
is obtained by setting $V_i=0$. Selfinteractions $V_i S_i$ 
determined by the the linear response method have also been 
introduced in \cite{KaRo99} as a heuristics
to correct the naive MF equations for Boltzmann machines.
A sanity check of the internal consistency of our approach 
is obtained by the fact that the matrix $\chib$ must be positive
definite. (If a group of the variables are complex, 
this has to hold for the submatrix of the real random
variables).

The next task is to compute the adaptive   
TAP approximation to the free energy $F(\Jb, \tb) = -\ln Z(\Jb, \tb)$. 
It is useful to generalize our model eq.\ (\ref{model}) 
to a one parameter class of models
where the interaction $\Jb$ is replaced by $s\Jb$ with 
$0\leq s\leq 1$ and to define the 
Legendre transform (Gibbs free energy) by
$$
\Phi_s(\mb,\Mb)= 
F(s \Jb +\lab,\tb +\gab) + \sum_i \gamma_i m_i   
+ \sum_i \frac{\la_i}{2} M_i,
$$
where $\gamma_i$ and $\lambda_i$ are external fields 
conjugate to  $S_i$ and $S_i^2$ which must be chosen 
to extremize the right hand side and
$\lab$ is a diagonal matrix with entries $\lambda_i$.
The solutions $\mb^e$ and $\Mb^e$ of the sets of equations
$
\partial_{m_i} \Phi_s = \partial_{M_i} \Phi =0 ,
$
determine the 
correct equilibrium expectation values $\<S_i\>_s = m_i^e$ and
$\<S_i^2\>_s = M_i^e$ (the index indicates that 
the expectation is taken with parameter $s$). 
Our desired approximation to the free energy is finally obtained 
as $F(\Jb, \tb) = \Phi_1(\mb^e,\Mb^e)$. To compute $\Phi_1$ we differentiate
$\Phi_s$ with respect to $s$, to show that 
\beq
\Phi_1 = \Phi_0 - \oneh\int_0^1 ds\; \left\{ 
\sum_{i,j} m_i J_{ij} m_j + \tr(\chib_s \Jb) \right\}
\eeq
with $\chi_{s,ij} = \<S_i S_j\>_s - \<S_i\>_s \<S_j\>_s$. 
Inserting our TAP approximation $\chib_s = (\Lab_s - s\Jb)^{-1}$
and integrating, we obtain
\beqa \label{generalG}
\Phi_1 &  = & \Phi_0 -  \oneh \sum_{ij} m_i J_{ij} m_j + \Delta \Phi \\
\Delta \Phi & = &
\oneh \ln \det (\Lab - \Jb) - \oneh \sum_i V_i \chi_{ii}
+ \oneh \sum_i \ln \chi_{ii}  \nonumber 
\eeqa
with $\chi_{ii} = M_i - m_i^2$.
The first two terms constitute the naive mean field approximation to $\Phi$
and the last term $\Delta \Phi$ is the Onsager correction.
Note, that this result is not equivalent to a truncation of a
power series expansion of $\Phi$ to second order in $s$ (a Plefka expansion
\cite{Pl82}) but 
%(except for the SK model) 
contains terms of all orders.
A different way to derive this result 
is obtained from the observation that the functional form of the 
Onsager term $V_i$ in the TAP equations 
does not depend on the specific single variable densities $\rho(\Sb)$.
Hence, we may compute this universal form  
by calculating $\Phi$ for an exactly solvable model, i.e.\ for a
Gaussian $\rho$ and subtract the naive mean field part.
This is related to the strategy used by Parisi and Potters \cite{PaPo95} 
in order
to derive the TAP equations for a spin glass model with orthogonal
random matrix $\Jb$. 
%In our case, the result is a consequence of the cavity derivation.

To check the significance of our approach, we will next show
that it will give the correct results for the statistical mechanics 
in the thermodynamic limit 
$N\to\infty$ for a large class of distributions of the random matrix $\Jb$.
For simplicity, we specialize to models with only one type of
single variable distribution $\rho_i(S) = \rho(S)$. 
\cut{Generalizations 
will be given elsewhere \cite{OpWi00b}.} 
Selfaveraging properties of the models
can be computed within the replica framework by averaging the free energy
over the distribution of the random matrix $\Jb$. This 
requires the calculation of the asymptotic scaling of the function
$
K_N (\Ab) \equiv \frac{1}{N}\ln \left[e^{\oneh\tr (\Ab \Jb)}\right]_{\Jb} 
$
for the matrix $A_{ij} = \sum_{a=1}^n S_{ia} S_{ja}$, where the $\Sb_a$,
are $n$ replicas of the variables. Following Ref.\ \cite{PaPo95} and assuming
the scaling $K_N (\Ab) \simeq \tr G(\Ab/N)$ as $N\to\infty$ where the function
$G$ characterizes the random matrix ensemble, the averaged free energy 
will depend only on the {\em single} set of orderparameters given by
$q_{ab}\equiv \frac{1}{N} \sum_i S_{ia} S_{ib}$. This is characteristic
for models with matrices $\Jb$ of extensive connectivity. 
\iffalse
For those, we expect 
that our assumption of weak dependencies are fulfilled. It will also
{\em exclude} models with {\em finite} connectivities for which  
an infinite number of multi-spin orderparameters will appear.
\fi
E.g., the SK-model with coupling matrix of independent components of variance 
$\frac{\beta}{N}$ has $G(r)= \frac{(\beta r)^2}{4}$ and 
the Hopfield model with
$J_{ij}=\sum_{\mu=1}^{\alpha N} x_i^{\mu} x_j^{\mu}$ and independent
$x_i^{\mu}$ with variance 
$\frac{\beta}{N}$ leads to
$G(r)= -\frac{\alpha}{2}(\ln(1-\beta r)+ \beta r)$.  
\iffalse
Models of learning from teacher data are incorporated in this framework
by a Bayesian integration over the teacher which restores the symmetry that
would be broken by the occurrence of a single teacher. 
\fi
Under the assumption of replica symmetry, 
the averaged free energy $f= - \frac{1}{N} [\ln Z]_{\Jb}$ is
obtained by extremizing
\beqa\label{F:rep}
f(q,\D) & = &  -G(\D) + \D \left(q G{''}(\D) + G'(\D)\right)  \\
- \int Dz \ln  \int & dS &\rho(S) 
\exp\left[ \sqrt{2 q G{''}(\D)} z S  + G'(\D) S^2 \right] 
\nonumber
\eeqa
with respect to the off-diagonal orderparameter $q = q_{ab}$
and to $\Delta = q_{aa} -q$, where $Dz = \frac{dz}{\sqrt{2\pi}}\; e^{-z^2/2}$.

We can show the correspondence for $N\to \infty$ 
of the adaptive TAP method and replica theory. 
A disorder average gives
the conventional TAP result for the Onsager coefficients: 
$
V_i = V = 2 G'(\overline{\chi})\ ,
$
with $\overline{\chi} = \frac{1}{N} \sum_i [\chi_{ii}]_{\Jb}$. 
To compare the TAP Gibbs free energy eq.\ (\ref{generalG})
with the replica symmetric free energy (\ref{F:rep}), we compute
$
\hat{f}= - \lim_{N,\gamma \to\infty} \frac{1}{\gamma N} \left[ 
\ln \int d\mb\; d\Mb 
\exp(-\gamma \Phi) \right]_{\Jb},
$
where the
paths of integration must be chosen such that the integral converges.
The integral will be dominated by the values
for $\mb$ and $\Mb$ which fulfill the TAP equations. 
Evaluating this expression using the replica method shows that 
both free energies coincide, i.e. $\hat{f} = f$.
It is also possible to translate the condition of positive definiteness 
of the susceptibility matrix $\chib$ into the thermodynamic limit. We can show 
that this stability is satisfied for
$
1 -  2 G''(\overline{\chi}) \frac{1}{N} \sum_i [\chi_{ii}]^2_{\Jb} > 0,
$ 
which coincides with the well known 
AT stability condition of replica theory \cite{MePaVi87}.

\begin{figure}[ht]
\vspace{6.7cm}
\includegraphics{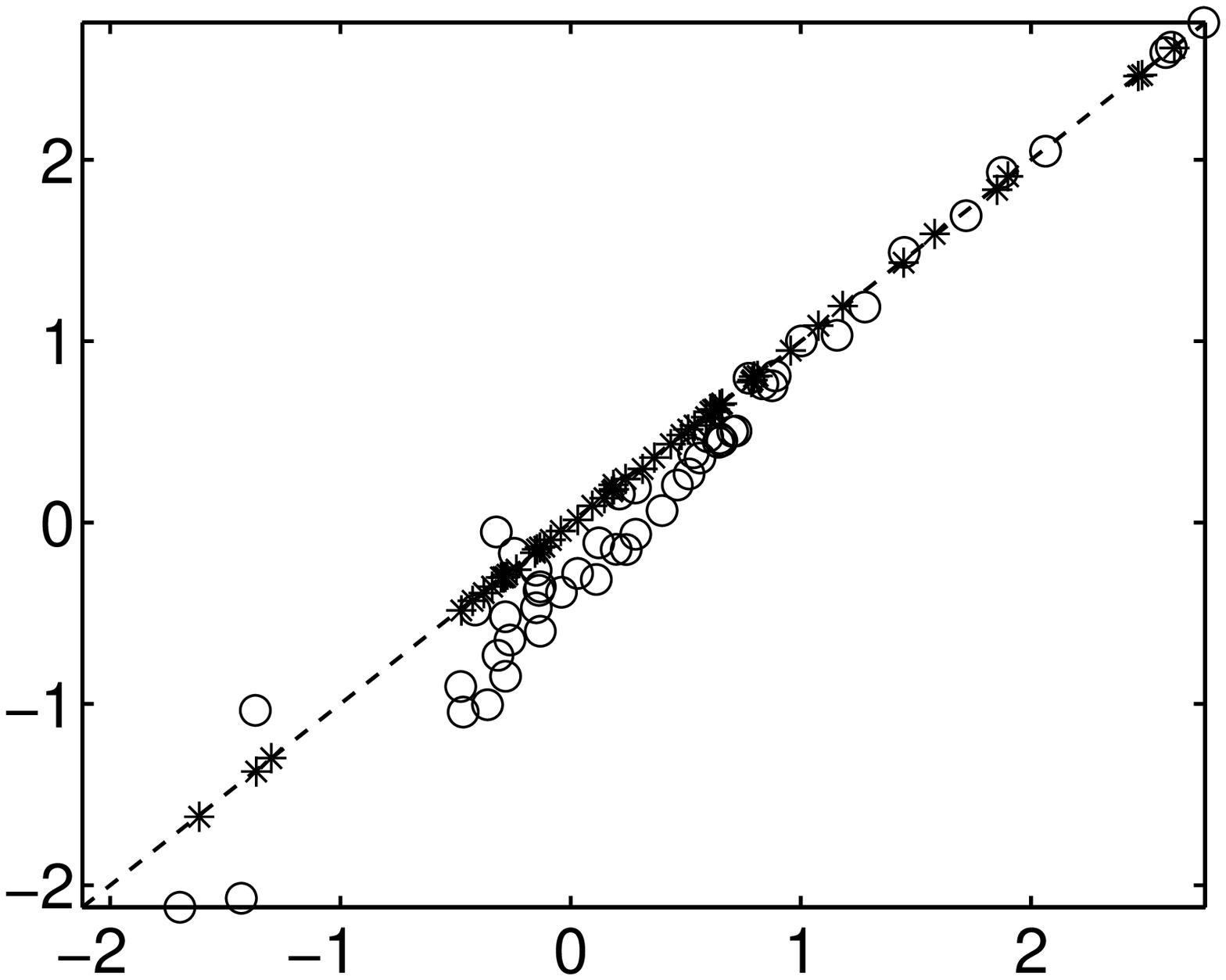}
\includegraphics{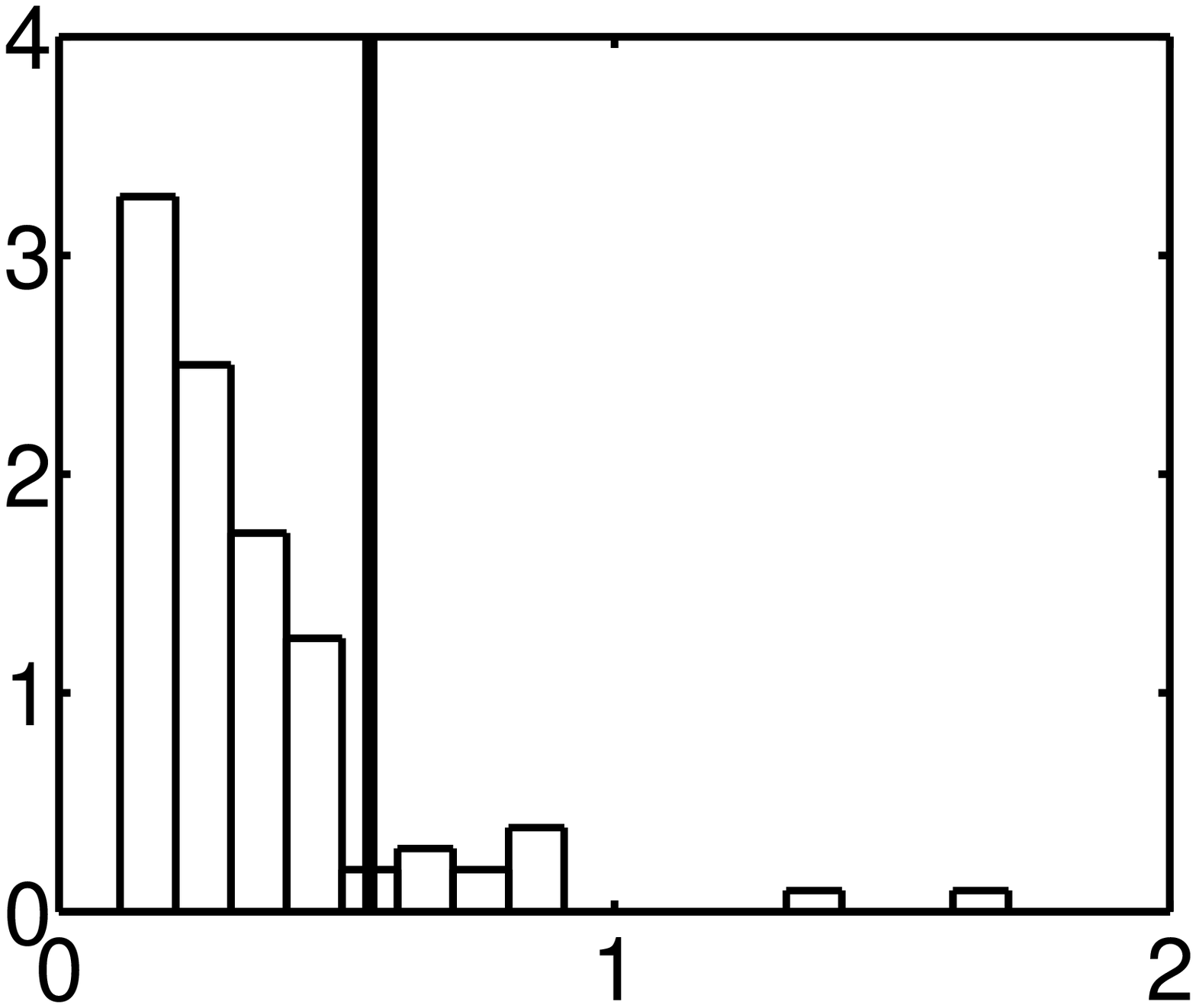} 
\begin{picture}(0,0)(-10,-68)
\put(-9,40){$y_j \< \hat{h}_j \>_{\backslash j}$}
\put(137,-65){$y_j \< \hat{h}_j \>_{\backslash j}^{\rm exact}$}
\put(192,12){{\large $P(V)$}}
\end{picture}
\caption{Test of self--consistency of 
TAP -- $y_j \< \hat{h}_j \>_{\backslash j}$ 
versus $y_j \< \hat{h}_j \>_{\backslash j}^{\rm exact}$. 
The stars/circles are for adaptive/conventional TAP. 
The inset shows the distribution of $V_i$ with the thick line indicating the 
conventional TAP solution.}
\label{comparisonSonar}
\end{figure}
We have performed two types of simulations of the TAP approaches on 
Bayesian neural network learning problems. 
For the first case (Fig.\ \ref{comparisonSonar}) we 
test the self--consistency of our method on a real  
data set, `Sonar -- Mines versus Rocks' \cite{GoSe88} of size $m=104$
with with binary class labels $y_j=\pm1$ and a $N=60$ dimensional input space. 
The prior is
$P_0(\Sb)\propto \exp (- \Sb \cdot \Sb/2)$ and
the likelihood $F(\hat{h}_j)=\phi( y_i\hat{h}_j/\sigma)$, with 
$\hat{h}_j=\sum_i x_{ij} S_i$,
$\phi(t)=\int^{\infty}_{-t} Dz$ and $\sigma^2=0.5$. 
We compute the prediction for the average (conjugate) cavity field  
$\< \hat{h}_j \>_{\backslash j}=\< \hat{h}_j \>
 - \hat{V}_j \< \hat{S}_j \>$, using eq. (\ref{cavfield}). 
The fraction of negative terms 
$y_j\< \hat{h}_j \>_{\backslash j}$ 
equals the `leave-one-out' estimate $\epsilon_{\rm loo}$ 
which provides an important 
practical estimator for the generalization error of the network.
If our theory takes the reaction of the remaining variables correctly 
into account, this prediction should be close to the `exact' 
average cavity field obtained by leaving one example out 
and solving the TAP eqs.\ for the remaining $m-1$ examples.
Fig.\ \ref{comparisonSonar} shows excellent agreement 
between the two computations and we find 
$\epsilon_{\rm loo}=\epsilon_{\rm loo}^{\rm exact}=33/104$. 
For comparison, the conventional TAP approach \cite{OpWi96}, 
which assumes a distribution of input data vectors 
with independent components, leads to a
wrong result, 
\iffalse
 performs significantly
worse. Hence, the computation of the 
leave-one-out estimate based on eq. (\ref{cavfield})
for this simpler TAP method leads to wrong results 
\fi
$\epsilon_{\rm loo}=41/104$ and $\epsilon_{\rm loo}^{\rm exact}=33/104$. 
\begin{figure}[ht]
\vspace{6.7cm}
\includegraphics{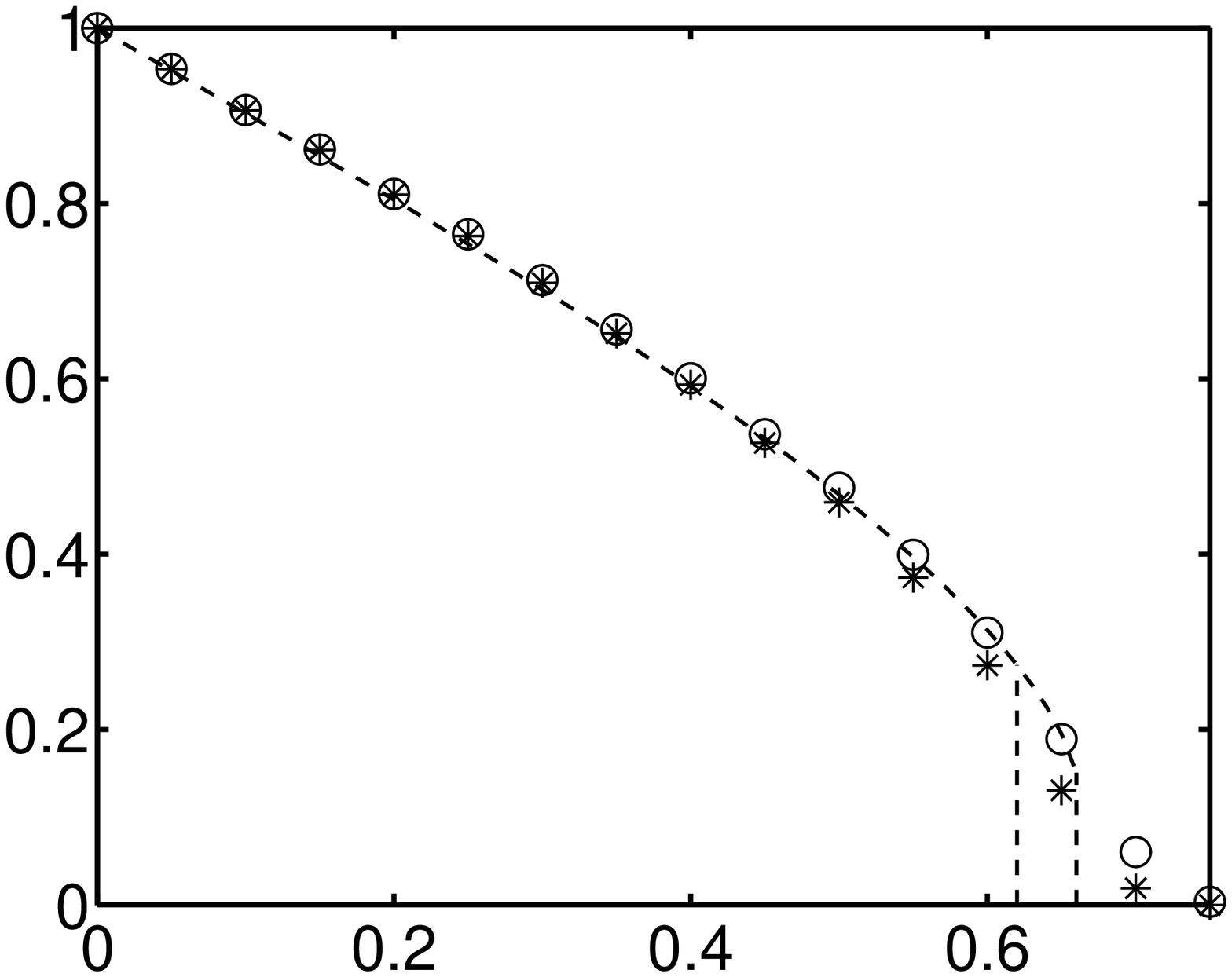}
\includegraphics{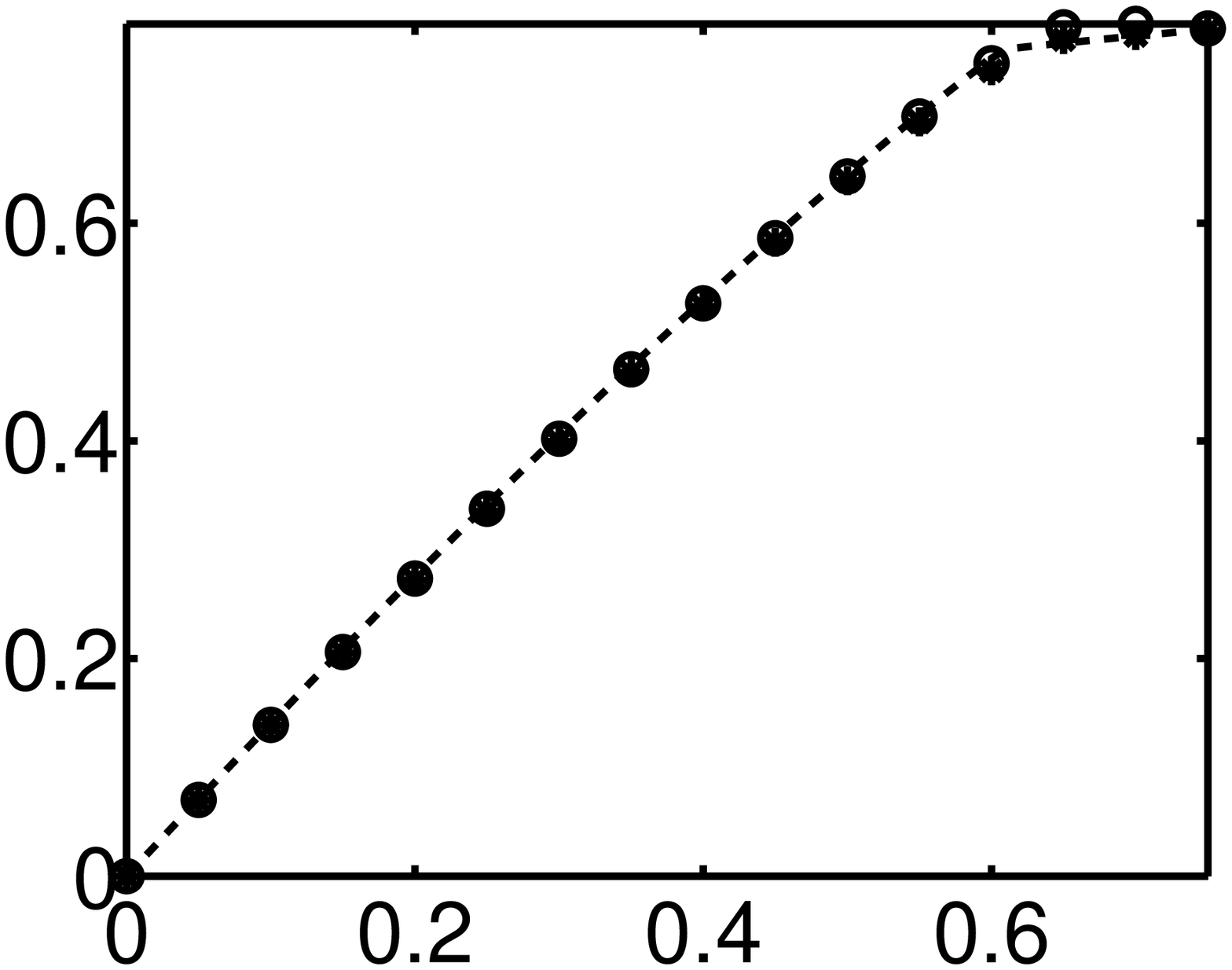} 
\begin{picture}(0,0)(-10,-68)
\put(-3,32){{\LARGE $\epsilon$}}
\put(149,-60){{\LARGE $\frac{m}{N}$}}
\put(105,17){{\LARGE $\frac{\Phi}{N}$}}
\end{picture}
\caption{Learning curve for the linear Ising perceptron -- test error rate 
against number of training examples. 
The stars/circles are for adaptive/conventional TAP.
The dashed lines are the replica result where the first vertical is 
the thermodynamic transition and the second the spinodal point where the 
meta-stable solution vanishes. The inset shows the corresponding 
normalized free energy $\Phi/N$. The simulations are averaged over 100
runs with error bars of the size of the symbols.}
\label{linearIsing}
\end{figure}

In the second set of simulations Fig.\ \ref{linearIsing} we demonstrate    
that the adaptive TAP method yields the correct statistical physics 
for the case of the linear Ising perceptron
\cite{SeSoTi92}. This has prior distribution
$P(S)= \frac{1}{2} \delta(S-1) + \frac{1}{2} \delta(S+1)$
and likelihood $F(\hat{h}_j) \propto \exp(-(y_i-\hat{h}_j)^2/2\sigma^2)$,
where we have chosen $\sigma^2=0.2$ and $N=60$ in the simulations. 
See Ref.\ \cite{Tanaka2000} for a discussion of this model in the context 
of demodulation in communications systems. 
\iffalse
This system has
also been discussed recently as a model of a multi--user demodulator in 
the so-called {\em Code-Division-Multiple-Access} technique for 
communications systems \cite{Tanaka2000}.
\fi
To compare with the replica results \cite{SeSoTi92}, we have generated
inputs at random and outputs using a noise free teacher perceptron
sampled from the same prior.  The small deviations between theory
and TAP simulations close to the 
first order transition are attributed to hysteresis effects.

It will be interesting to extend our adaptive TAP method to
glassy systems (generalizing the ideas of chapter V in \cite{MePaVi87})
where the present approach would fail, e.g. 
indicated by the appearance of negative eigenvalues in the
susceptibility matrix. However, one may speculate that in such cases 
solving the TAP equations may be highly nontrivial.
{\it Acknowledgments.} 
We thank T. Tanaka for providing us with his  
preprint \cite{Tanaka2000}. 
This research is supported by the Swedish
Foundation for Strategic Research as well as the Danish Research
Councils through the Computational Neural Network Center (CONNECT) and
the THOR Center for Neuroinformatics.
%\begin{thebibliography}{99}

%\end{thebibliography}

\end{document}